\documentstyle {article}

\def\L{{\cal L}}
\def\nn{\nonumber}
\def\ll{\lambda}
\def\sp{\sqrt{p_2}}
\def\N{\newline}
\def\o{\over}
\def\p{\partial}
\def\l{\label}
\def\z{\zeta}
\def\be{\begin{equation}}
\def\bea{\begin{eqnarray}}
\def\ee{\end{equation}}
\def\eea{\end{eqnarray}}
\begin{document}
\begin{titlepage}
\begin{center}
\vspace*{1.5cm}
\hfill
\vbox{
    \halign{#\hfil         \cr
           hep-th/9708073 \cr
           IPM-97-226\cr
           August 1997\cr} 
      }  
\vskip 1cm
{\large \bf
On the Picard-Fuchs Equations of $N=2$ Supersymmetric $E_6$
Yang-Mills Theory}
\vskip .6in
{\bf A.M. Ghezelbash$^{*,\dagger,}$\footnote{e-mail:amasoud@physics.ipm.ac.ir},
A. Shafiekhani$^{*,}$\footnote{e-mail:ashafie@theory.ipm.ac.ir},
M.R. Abolhasani$^{*,+,}$ \footnote{e-mail:abolsan@physics.ipm.ac.ir
}}\\
\vskip .25in
{\em
$^*$Institute for Studies in Theoretical Physics and Mathematics, \\
P.O. Box 19395-5531, Tehran, Iran.\\
$^\dagger$Department of Physics, Alzahra University, Vanak,
Tehran 19834, Iran.\\
$^+$Department of Physics, Sharif University of Technology,
P. O. Box 19365-9161, Tehran, Iran.}
\end{center}
\vskip 1.5cm
\begin{abstract}
We obtain the Picard-Fuchs equations of $N=2$ supersymmetric Yang-Mills
theory with the exceptional gauge group $E_6$. Such equations are based on
$E_6$ spectral curve.
\end{abstract}
\end{titlepage}\newpage
In the last few years, enormous advances have been made in understanding
of the low-energy behaviour of $N=2$ supersymmetric gauge theories. The
progress was initiated with the paper of Seiberg and Witten \cite{1},
where the exact low-energy Wilsonian
effective action of the pure $N=2$ supersymmetric Yang-Mills theory with
the gauge group $SU(2)$ was derived. Since then, their work has been
generalized to the supersymmetric pure gauge theories with
other gauge groups \cite{2} and to the theories with the matter multiplets
\cite{3}.
In principle, the exact solution of such theory is given by an algebraic
curve. In the case of theories with classical Lie gauge groups, the algebraic
curves is hyperelliptic \cite{4} which of course should satisfy a set of
consistency conditions. The hyperelliptic curve of
the theories with the exceptional gauge groups are constructed in \cite{5}.

To understand the strong coupling region of the theory, one defines
the Higgs fields $\vec a$ and their duals $\vec a_{D}$
which are related to the prepotential of the low-energy effective action,
$F(\vec a)$, by $\vec a_{D}={{\p F}\o {\p \vec a}}$.
These fields which are periods of the Riemann surface
defined by the given algebraic curve, are represented by
the contour integrals of the Seiberg-Witten differential one-form
$\ll$,
\be \l{AAD}
\vec a=\int _{\vec \alpha}\ll,\qquad\vec a_{D}=\int _{\vec \beta}\ll ,
\ee
where $\vec \alpha$ and $\vec \beta$ are homology cycles on the Riemann
surface.

To obtain the periods, $\vec a$ and $\vec a_D$, one can derive the
Picard-Fuchs (PF) operators, which annihilate $\vec a$ and $\vec a_{D}$.
The PF equations have been derived in the case of theories with the
pure classical gauge groups and also classical gauge groups with
massless and massive multiplets \cite{6}.
In all these cases, the underlying algebraic curve of the theory is a
hyperelliptic curve.\newline
On the other hand in \cite{8}, the algebraic curve of the supersymmetric
gauge theory is constructed from the spectral curve of the periodic Toda
lattice.
In the case of theories with the classical gauge groups,
these curves are equivalent to the
hyperelliptic curves.
But recently in \cite{9}, the PF equations of the supersymmetric $G_2$
gauge group have been constructed from the spectral curve of the
$(G_2^{(1)})^{\vee}$ Toda lattice theory. Also it has been shown that the
calculation of $n-$instanton effects  agrees with the microscopic results,
while the hyperelliptic curve shows different behaviour
compared with the microscopic results of \cite{10}.

Our motivation in this letter, is to shed
light onto the strong coupling behaviour of $E_6$ theory and the comparison of
the spectral curve of $E_6$ theory with the hyperelliptic curve of the same
theory.
As a first step, we derive explicitly a set of PF equations of the
theory
with $E_6$ gauge group. In next step, we will give solutions of these
equations, and the multi-instanton corrections to the prepotential
of $E_6$ theory \cite{105}.

We use the spectral curve of $E_6$ given in \cite{11},
\be \l{YE6}
\zeta+{w\o \zeta}=-u_6+{{q_1+p_1\sqrt{p_2}}\o{x^3}},
\ee
which is obtained by degeneration of $K_3$ surface to an $E_6$ type
singularity. The polynomials $q_1$,$p_1$ and $p_2$ are given by,
\bea \l{Q1P1P2}
q_1&=&270 x^{15} + 342 u_1 x^{13} + 162 u_1^2 x^{11} - 252 u_2 x^{10} + (26
u_1^3 + 18 u_3) x^9\nn\\
&-& 162 u_1 u_2 x^8 + (6 u_1 u_3 - 27 u_4 ) x^7-(30 u_1^2 u_2 - 36 u_5 ) x^6
\nn\\
&+& (27 u_2^2 - 9 u_1 u_4 ) x^5 - (3 u_2 u_3 - 6 u_1 u_5 ) x^4 - 3 u_1 u_2^2
x^3
\nn\\
&-& 3 u_2 u_5 x -u_2^3;\nn\\
p_1&=&78 x^{10}+60 u_1 x^8 +14 u_1^2 x^6-33 u_2 x^5+2 u_3 x^4-5 u_1 u_2 x^3
\nn\\
&-& u_4 x^2 - u_5 x - u_2^2;\nn\\
p_2&=&12 x^{10}+12 u_1 x^8 +4 u_1^2 x^6-12 u_2 x^5+ u_3 x^4 - 4 u_1 u_2 x^3
\nn\\
&-&2 u_4 x^2 + 4 u_5 x + u_2^2,
\eea
where $u_1\equiv c_2,\,u_2\equiv c_5,\,u_3\equiv c_6,\,u_4\equiv c_8,\,u_5
\equiv c_9,\,u_6\equiv c_{12}$ and $c_i$ is $i$th Casimir of $E_6$ with weight
$i$.
By introducing the new variable $y=-\zeta - {w\o \zeta}$, the Seiberg-Witten
differential which is given by
$\ll =-2 x {{d\zeta}\o{\zeta }}$
becomes,
\be \l{LANDA} \ll ={{xdy}\o{\sqrt{y^2-4w}}}.\ee
The derivatives of $\ll$ with respect to the Casimirs of the group are
given by,
\bea \l{DER}
\p _i \ll ={{\p \ll}\o {\p u_i}}&=&-{{\p _i y}\o {\sqrt{y^2-4w}}}+d(*),\nn\\
\p _{ij}\ll={{\p ^2 \ll}\o {\p u_i\p u_j}}&=&{{\p _i\p _j y}\o {\sqrt{
y^2-4w}}}-{{y\p _i y\p _j y}\o{(y^2-4w)^{3/2}}}+d(*).
\eea
Now, we take the following form for the PF operator,
\be \l{PFFORM}
\L=a_{ij}(u_1,\cdots ,u_6)\p _{ij}+a_{i}(u_1,\cdots ,u_6)\p _{i}+a.
\ee
After applying $\L$ to $\ll$ given in (\ref {LANDA}), we get,
\bea \l{LLANDA}
\L\ll&=&{{L_1}\o {\z }}+{{L_2}\o {x^2\z }}+{{L_3}\o {x^3\z }}+{{L_4}\o {x^2\z
\sp }}+{{L_5}\o {x^3 \z \sp}}+
{{L_6}\o {x^3 \z p_2 \sp }}+{{L_7}\o {\z ^{3}}}
+{{L_8}\o {x^3\z ^3}}\nn\\&+&{{L_9}\o {x^3\z ^3 \sp }}+{{L_{10}}\o {x^6\z ^3}}+
{{L_{11}}\o {x^6\z ^3 \sp }}+{{L_{12}}\o {x^6\z ^3 p_2}}+{{L_{13}}\o {x^9\z ^3
}}+
{{L_{14}}\o {x^9 \z ^3 \sp}}+{{L_{15}}\o {x^9 \z ^3 p_2}},
\eea
which $L_1,\cdots ,L_{15}$ are complicated polynomials of $x,
u_1,\cdots,u_6$ and $w$. The form of eq. (\ref{LLANDA}) hints to choose
the following form for $\L\ll$,
\be \l{DFG}
\L\ll=d({{f(x)}\o{x^2\z \sp}}+{{g(x)}\o{x^2\z}}).
\ee
To find coefficients $a_{ij}$, $a_i$ and $a$ in (\ref{PFFORM}),
we write the following expansion of these coefficients according to their
weights, such that the weight of $\L$ becomes zero,
\bea \l{AAA}
a_{ij}&=&\sum _{p=1}^{N_{ij}} \bigg (d_{ij,p}\prod _{k=1}^7 u_k^{n_k}\bigg ),
\nn\\
a_{i}&=&\sum _{p=1}^{N_{i}} \bigg (d_{i,p}\prod _{k=1}^7 u_k^{n_k}\bigg ),
\eea
where $N_{ij}$ and $N_i$ are the number of different solutions of the
relations
$\{\sum _{k}n_k[k]=[i]+[j]\}$
and
$\{\sum _{k}n_k[k]=[i]\}$ respectively, and $[k]$ is the weight of $u_k$.
In eqs. (\ref{AAA}), $d_{ij,p}$ and $d_{i,p}$ are zero weight constants and
$u_7\equiv w$ has weight $24$.
Choosing the weight of $x$ equal to one, eqs. (\ref{LANDA})
and (\ref{DFG}) imply that $\ll$ has
weight one. Hence, $f(x)$ and $g(x)$ are polynomials of $x$ with
degrees $20$ and $15$
respectively. By these considerations, the general expression of $f(x)$
and $g(x)$ are as follows,
\bea \l{F}
f(x)&=&\sum _{i=0}^{20}\bigg (\sum _{j=1}^Nf_{i,j}\prod_{k=1}^7 u_k^{n_k}
\bigg ) x^i,
\\ \l{G}
g(x)&=&\sum _{i=0}^{15}\bigg (\sum _{j=1}^Ng_{i,j}\prod_{k=1}^7 u_k^{n_k}
\bigg ) x^i,
\eea
where $N$ is the number of
different solutions of $\{\sum _{k}n_k[k]+i=20\}$ in (\ref{F}) and
$\{\sum _{k}n_k[k]+i=15\}$ in (\ref{G}). By substitution of (\ref{AAA}),(\ref
{F}) and (\ref{G}) in (\ref{DFG}) and after some simplifications, we obtain
a polynomial of degree $120$ in $x$ that identically must be zero in $x$
and $u_i$'s.
Then we obtain a set of nonlinear equations for $333$ parameters,
$d_{i,p},\,d_{ij,p},\,a,f_{i,j},\,g_{i,j}$. Solutions
of these equations give a set of different PF operators.
As an example,
\be
\L _1=(u_1u_5+u_2u_3)\p _1\p _5+18u_2u_3u_5\p _4\p _6+6u_2^2u_3\p _4^2-4u_1u_5
\p _2\p _3,
\ee
which corresponds to the following $f(x)$ and $g(x)$,
\bea
f(x)&=&-(36u_1u_5+18u_2u_3)x^9-(18u_1^2u_5+12u_1u_2u_3)x^7+(9u_2u_1u_5-
3u_2^2u_3)
x^4,\nn\\
g(x)&=&-(9u_1u_5+3u_2u_3)x^4.
\eea
Other operators are as follows,
\bea
\L _2&=&-8u_1^2u_4\p _3^2-2u_1u_2u_4\p _3\p _5+3u_2^2u_3\p _4^2+u_2^2\p _2^2-12
u_2^2\p _1\p _4-6u_1u_4^2\p _3\p _6-18 u_1^2u_2^2\p _1\p _6-18u_1u_2^3
\p _2\p _6\nn\\
&-&u_1
u_4\p _1\p _4-18u_1u_2^2u_4\p _4\p _6-18u_1u_2^2\p _6;\nn\\
\L _3&=&u_6u_2\p _4\p _5+3u_2u_4^2\p _5\p _6+2u_2u_4\p _2\p _4
+4u_1u_2u_4\p _3\p _5+u_2^2u_4\p _5^2
-3u_6u_2\p _2\p _6;\nn\\
\L _4&=&12u_1^2u_2^2\p _3\p _4-u_1u_2\p _1\p _2+9u_1u_2^3\p _2\p _6
+36u_1^2u_2u_5\p _3\p _6-2u_1u_2u_4\p _3\p _5+9u_1u_2^3u_5\p _5\p _6
+9u_1u_2^2\p _6;\nn\\
\L _5&=&
8u_1^4\p _1\p _3+u_1^3u_2\p _1\p _5+8u_1^3u_4\p _3\p _4+12u_1^3
\p _3+4u_1^2\p _1^2-u_1^2u_3\p _1\p _4+2u_1^2u_5\p _2\p _4-u_1^2u_2u_4\p _4
\p _5\nn\\
&+&4u_3u_4\p _3\p _4-3u_3u_4\p _1\p _6-u_1^2u_2^2\p _2\p _5+3u_1^2u_4\p _6;\nn\\
\L _6&=&9u_2u_3u_5\p _4\p _6+2u_3u_5\p _3\p _5+(27u_3u_5^2+9u_1^2u_4u_6)\p _6^2
+2u_1^2u_6\p _4^2
+3u_1^2u_2u_6\p _5\p _6+12u_1^3u_6\p _3\p _6,
\eea
where the corresponding functions $f(x)$ and $g(x)$ are of degrees $(10,\, 5)$,
$(7,\, 2)$, $(13,\, 8)$, $(16,\, 11)$ and $(5,\, 0)$ respectively.
\vskip 1 cm
{\large CONCLUSIONS}
\newline
In this paper, by a systematic method, we obtain the PF equations of $N=2$
supersymmetric $E_6$ Yang-Mills theory. These equations are useful to
obtain the periods and multi-instanton corrections to the
prepotential of the theory which is based on $E_6$ spectral curve \cite{105}.
\vskip 1 cm
{\large ACKNOWLEDGEMENTS}
\newline
We would like thank to F. Ardalan for continuous encouragement.\N
After completion of this work, we received the paper \cite{ITO} which has
some overlap with this work.


\begin{thebibliography}{99}
\bibitem{1}
N. Seiberg and E. Witten, Nucl. Phys. {\bf B426} (1994) 19; ibid. {\bf
B431} (1994) 484.
\bibitem{2}
A. Klemm, W. Lerche, S. Yankielowicz and S. Theisen,
Phys. Lett. {\bf B344} (1995) 169;\N
A. Klemm, W. Lerche, and S. Theisen,
Int. J. Mod. Phys. {\bf A11} (1996) 1929;\N
P. Argyres, A. Faraggi, Phys. Rev. Lett. 73 (1995) 3931;\N
U. H. Danielsson, B. Sundborg, Phys. Lett. {\bf B358} (1995) 273;\N
A. Brandhuber, K. Landsteiner, Phys. Lett. {\bf B358} (1995) 73;\N
P. C. Argyres, A. D. Shapere, Nucl. Phys. {\bf B461} (1996) 437;\N
U. H. Danielsson, B. Sundborg, Phys. Lett. {\bf B358} (1995) 273.
\bibitem{3}
A. Hanany and Y. Oz, Nucl. Phys. {\bf B452} (1995)283;\N
A. Hanany, Nucl. Phys. {\bf B466} (1996) 85.
\bibitem{4}
A. Klemm, W. Lerche, S. Yankielowicz and S. Theisen,
Phys. Lett. {\bf B344} (1995) 169.
\bibitem{5}
U. H. Danielsson, B. Sundborg, Phys. Lett. {\bf B370} (1996) 83;\N
M. Alishahiha, F. Ardalan, F. Mansouri, Phys. Lett. {\bf B381} (1996) 446;\N
M. R. Abolhasani, M. Alishahiha, A. M. Ghezelbash, Nucl. Phys. {\bf B480}
(1996) 279.
\bibitem{6}
M. Matone, Phys. Lett. {\bf B357} (1995) 342;\N
H. Ewen, K. Foerger and S. Theisen. Nucl. Phys. {\bf B485} (1997) 63;\N
J. M. Isidro, A. Mukherjee, J. P. Nunes and H.J. Schnitzer, Nucl. Phys.
{\bf B492}
(1997) 6474174.\N
M. Alishahiha, Phys. Lett. {\bf B398} (1997) 100.
\bibitem{8}
E. Martinec and N. P. Warner, Nucl. Phys. {\bf B459} (1996) 97.
\bibitem{9}
K. Ito, hep-th/9703180.
\bibitem{10}
K. Ito and N. Sasakura, Phys. Lett. {\bf B382} (1996) 95;\N
K. Ito and N. Sasakura, Nucl. Phys. {\bf B484} (1997) 141.
\bibitem{105}
A. M. Ghezelbash, Phys. Lett. {\bf B423} (1997) 87.
\bibitem{11}
W. Lerche and N. P. Warner, hep-th/9608183.
\bibitem{ITO}
K. Ito and S.-K. Yang, hep-th/9708017.
\end{thebibliography}
\end{document}